\documentclass[aps,preprint,groupedaddress,showpacs,floatfix]{revtex4}
\usepackage{graphics}
\usepackage{epsfig}
\begin{document}
\title{Microscopic model of diffusion limited aggregation and electrodeposition 
in the presence of levelling molecules}
\author{G.J. Ackland and E.S.Tweedie}
\affiliation{School of Physics, The University of
Edinburgh\\
Edinburgh, EH9 3JZ, Scotland, United Kingdom \\ gjackland @ ed.ac.uk} 
\date{\today}
 
\begin{abstract}
A microscopic model of the effect of unbinding in diffusion limited
aggregation based on a cellular automata approach is presented. The
geometry resembles electrochemical deposition - ``ions''
diffuse at random from the top of a container until encountering a
cluster in contact with the bottom, to which they stick.  The model
exhibits dendritic (fractal) growth in the diffusion limited case.
The addition of a field eliminates the fractal nature but the density
remains low. The addition of molecules which unbind atoms from
the aggregate transforms the deposit to a 100\% dense one (in 3D).
The molecules are remarkably adept at avoiding being trapped.
This mimics the effect of
so-called ``leveller'' molecules which are used in
electrochemical deposition. 
\end{abstract}

\maketitle


Roughening in electrochemical deposition is one of a class of growth
problems.  The two main theoretical methods of tackling the problem
are continuum and microscopic methods.  The continuum approach 
is typified by the KPZ equation\cite{KPZ} and its variants\cite{Amar,bales,Halsey2}. 
Microscopic methods fall broadly into two classes: Eden-type\cite{Eden1958,Halsey3} 
models which consider the deposit growing into a surrounding medium, and 
aggregation-type models\cite{Vicsek} which consider diffusing 
particles becoming attached to the growing deposit.

 This work
was motivated by aqueous electrodeposition where metal cations in
solution are driven by an external voltage to coat the cathode.
In practical electrodeposition work, for applications from
microelectronic interconnects to copper plating, the aim is usually to
avoid roughening and obtain a flat surface\cite{Iwamoto,Pastor}. 
Empirically, it has been known for many years that 
so-called ``levellers'' - typically organic molecules, can be added to
control the roughness of the films\cite{oniciu}.  The molecular-level
mechanism by which these levellers work remains uncertain, several
models have been proposed\cite{oniciu} including diffusion, chemical
filming, electrosorption, complex formation and ion-pairing. A continuum
model of the effect of levellers was recently advanced\cite{Haataja},
based on the assumption that the molecules were found preferentially
at high curvature regions, and have the effect of blocking further
deposition there.  It is likely that the dominant mechanism is
system-dependent.

The concentration of levellers in solution is typically a few orders 
of magnitude lower that of the metal ions,  and
the levellers are absorbed into the deposit in much lower concentration. 
This makes it unlikely that their effect results from strong
bonding of molecules to the surface.  This is borne out by {\it ab
initio} calculation of copper and organic molecules, which show that
phenyl rings are typically only weakly physisorbed to the surface of
the copper, but are strongly bound to single atoms or ions.
This arises because the metal atom/ion orbitals are able to lower
their energy by delocalizing and hybridising with the phenyl states,
however the metal surface states are already delocalized, and cannot gain
much energy by further hybridisation.\cite{emk}

In this paper we advance a microscopic picture based on this which
exhibits a levelling effect in diffusion limited growth.  The
mechanism is different from blocking - we postulate that the aromatic
molecules bind to metal ions and abstract them from the growing
deposit.  This simple mechanism gives some fascinating dynamics and 
provides an extremely strong levelling effect.

The model is based on the motion of autonomes\cite{autonome} on a lattice in two or three dimensions.
It is rule-based and comprises a regular lattice
occupied by  three types of autonome, ions in solution (I), 
deposited atoms (A) and molecules in solutions (M).  They move on a 
square (cubic) lattice according to the following iterated rules.

\begin{itemize}
\item 1/  Ions and molecules are introduced stochastically at the top, 
with probability $\mu_I$, $\mu_M$ (effectively a chemical potential), 
and may diffuse back out from the top.
\item 2/  All ions move stochastically in 2D(3D) in one of eight(26) directions, with a 
    bias in favour of moving downward (the field, $E$).
\item 3/  Molecules move stochastically in one of eight(26) directions. 
\item 4/  If the randomly chosen move would take the ion or molecule onto an 
    occupied space, no motion occurs.
\item 5/  Ions moving adjacent to the bottom, or a continuous chain of atoms 
connected  to the bottom are deposited (converted to atoms, which do not move)
\item 6/  Molecules adjacent to atoms convert the atoms into diffusing 
ions with "unsticking" probability $p$. 
\end{itemize}

The model depends on four parameters, the leveller and ion chemical
potentials $\mu_I$ and $\mu_M$, the unsticking probability $p$ and the
field $E$.

The simple case without levelling molecules is just diffusion limited
aggregation DLA\cite{Witten}.  This has been well studied\cite{Halsey}
and produces fractal geometries which can be found
experimentally\cite{Vasquez}.  To make contact with this work, it is
convenient to measure the density and/or fractal dimension of the
deposit rather than the surface roughness {\it per se}.

 The fractal nature is measured by\cite{Vicsek}

\begin{equation} \alpha=\lim_{x\rightarrow\infty}\frac{d\log(\rho(x))}{d\log{x}}.  
\end{equation}

where
$x$ is the height above the bottom and $\rho(x)$ is the fraction of cluster
atoms at that height.  Although the boundary conditions mean that the 
fractal nature is manifest only in
one direction, it is possible to define an effective fractal dimension
for the cluster as $D=d-\alpha$.\cite{Vicsek}

\begin{figure}\protect{\includegraphics[width=\columnwidth]{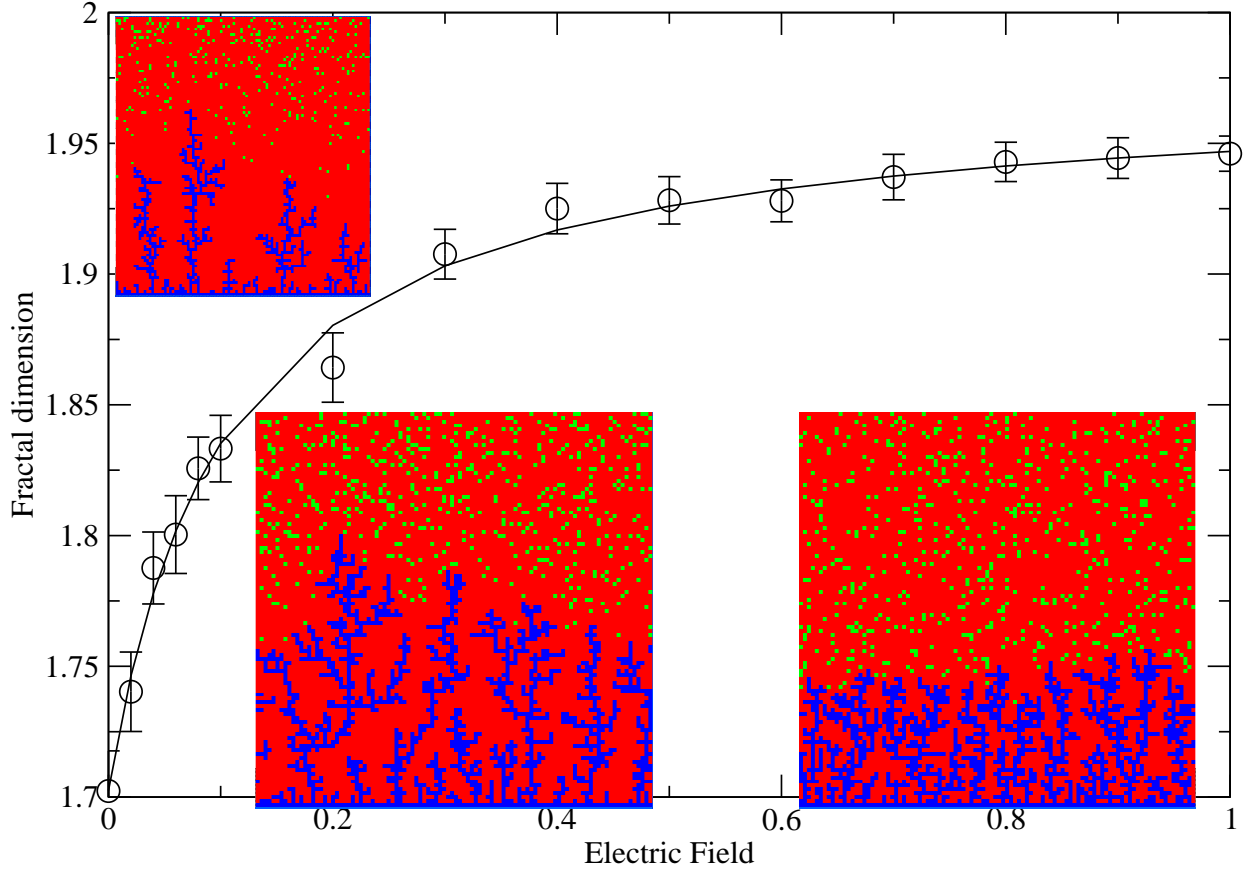}}
\caption{ Figure shows the effect of electric field 
on the fractal dimension of the 2D deposit without levellers.  
The field is applied as a bias between downward and upward hopping of
the ions - the downwards hops have probability (1+E)/8, sideways 1/8
and upwards (1-E)/8. 
Once the deposit reaches the top of the 
simulation cell, its growth is again affected by its inability to grow further.
Hence the density is measured from the central 80\% of the simulation, 
ignoring the upper and lower parts. 
We ran the simulation until the first "atom" reached the top,  
and measured the fractal dimension of the cluster and found that
calculations on a grid 500 sites wide and 100 high with periodic 
boundary conditions reliably reproduced this provided data above 
x=90 is discarded.  Indeed we were unable to determine a finite width effect on
$\alpha$ for calculations with widths down to 100. For increased height, 
the fractal dimension increases towards 2 for all non-zero fields.
Each datapoint was averaged over five runs to
obtain standard error bars.  In each run we first plot $\ln\rho$ (log
density) as a function of $\ln x$ (log height), $\alpha$ is the slope of
this graph: in practice, the termination of the simulation when
growing cluster reaches the top anywhere, causes a nonlinear tail to
this graph for large $x$, which is excluded from the fit.  Line is a
three-parameter fit $D=1.972-0.27/(1+9.696E))$. Upper insert shows 
fractal growth at zero field, lower inserts show snapshots
of the growing deposit at E=0.1 and E=0.5.
}\label{fig.DvsF}
\end{figure}

\begin{figure}\protect{\includegraphics[width=\columnwidth]{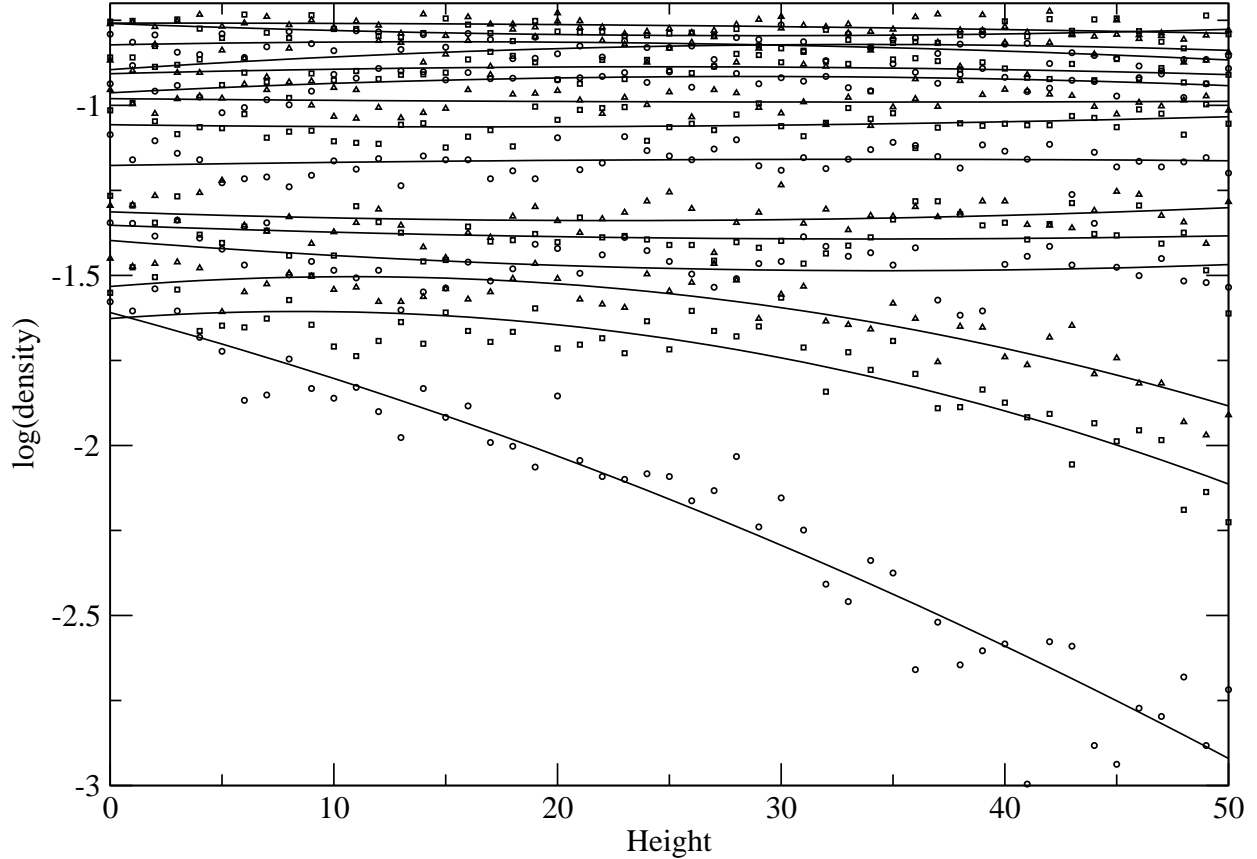}}
\caption{Plot of log density against height for 2D with various
applied fields ($\mu_M=0$, $\mu_I=0.05$, $E$ increment from 0-0.3 in steps of 0.02). Simulation size was 200x100, - the simulation was
stopped once any part the growing deposit reached y=100, at which
point the region above about y=50 is not in equilibrium. Symbols
(alternating circles, squares, triangles) represent actual densities
averaged over five samples); lines are quadratic fits to the data.  The
slope in the zero field case (lowest line) shows the fractal nature of
the growth, the pronounced curvature for fields of 0.02 and 0.04 shows
these are affected by finite size effects, the upper parts of the sampling region have not reached their equilibrium density.  The essentially straight
lines with zero slope for higher fields indicate non-fractal clusters,
while the value of the mean
density (0.3-0.5) shows the clusters are far from compact.  3D 
behaviour is qualitatively similar, with fractal behaviour observed only for E=0}\label{fig.DvsZ}
\end{figure}

When a field is added, the preferred direction breaks the
scale free fractal nature of the DLA deposit (see
fig. \ref{fig.DvsZ}. However, as can be seen in Figure \ref{fig.DvsF}
the deposit still has low density and a rough, dendritic surface.  The
same is true in the 3D case.  The profile of a typical simulation 
and its associated height-density relation is shown in Figure \ref{fig.DvsZ}.

The levelling molecules have the effect of unsticking atoms from the
deposit, thus the dendrite structure is cut away at its roots.  
This is a non-local process and evaluating the (dis)connected cluster is the 
most time-consuming computational aspect of the simulation.

\begin{figure}\protect{\includegraphics[width=\columnwidth]{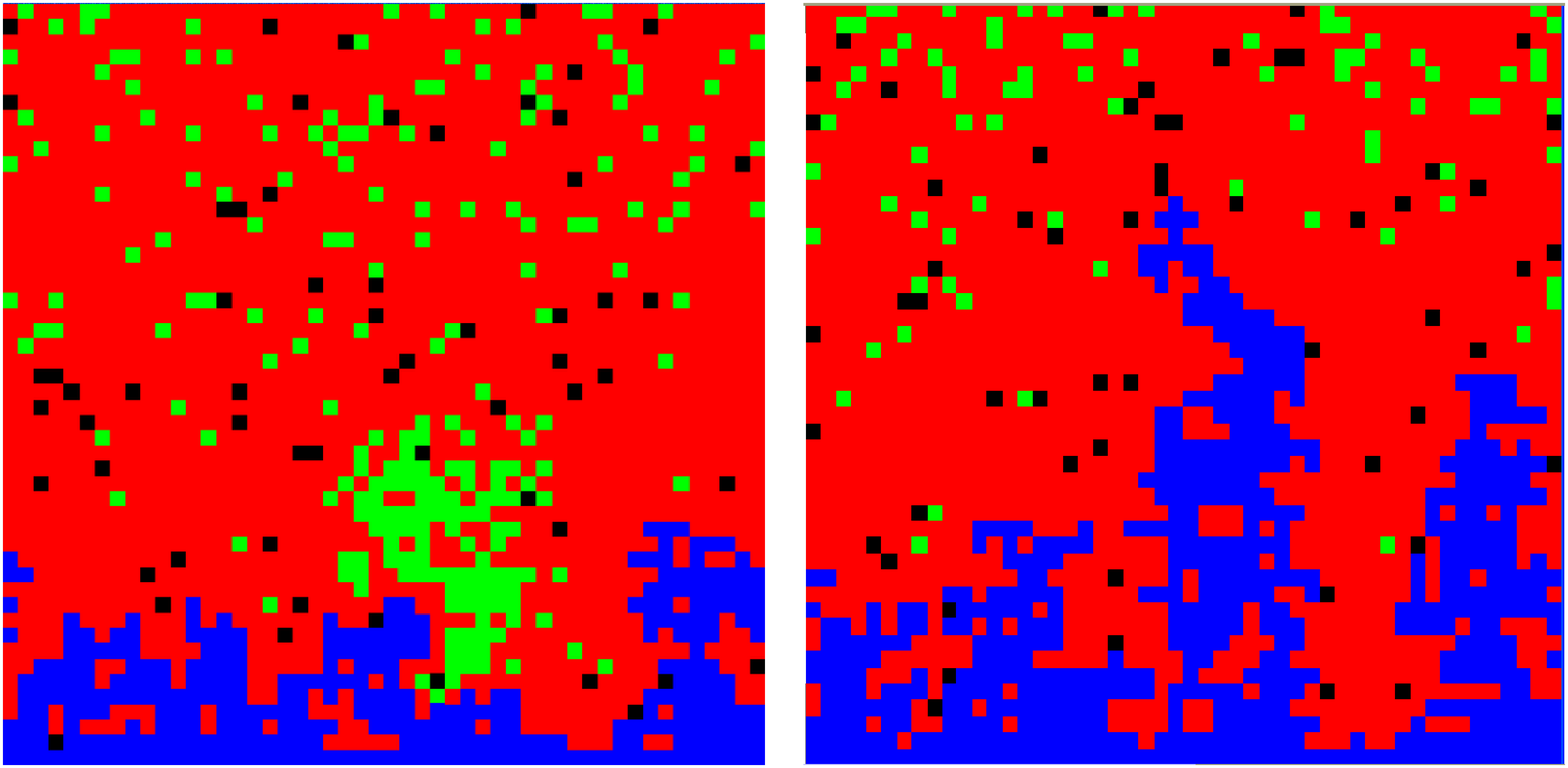}}
\caption{Snapshot of a levelling event.  In this
close-up of part of a 2D simulation, the black squares represent
molecules, the blue atoms and the green ions. (left) The large cluster of
green ions in the center of the figure have been disconnected by the
action of the single black molecule at their foot, and are about to fall. (right) some time later, the deposit has compacted, but the large central 
protruberance remains.
}\label{fig:undercut}
\end{figure}

First, we consider the effect of levellers in the two dimensional case.  Although not
applicable to real electrodeposition, the 2D case contains most of the
relevant physics. 
For 2D DLA in this geometry we obtain the expected $\alpha=0.29$\cite{Vicsek} 
(i.e. D=1.71),  levellers cause densification by undercutting the
fractal dendrites of the deposit.  Fig. \ref{fig:undercut}

Increasing the applied field has the effect of reducing $\alpha$ to a 
limiting value of about 0.05.  However, as can be seen in figure \ref{fig.DvsF}
eliminating the fractal growth does not lead to a high density deposit.

The addition of even a small amount of leveller with an unsticking
probability of 0.1 has a dramatic effect on the structure.  The 
fractal nature is lost, and the density of the
deposit is increased substantially - even a $\mu_M=0.01$ gives a dense
deposit.  The mechanism for this is that the levellers can diffuse
into the open deposit, and undercut the fingerlike growth.  Unsticking
a single atom can affect quite large numbers of atoms, by undercutting
structures bonded by a single connection (see figure \ref{fig:undercut})

Once leveller is present, 
increasing the unsticking probability above a rather small threshhold
value (0.05 for a $\mu_M=0.01$, $\mu_I=0.02$) does not give a
significant further increase in the density.  For high values of $p\mu_M$ the 
growth rate goes to zero: we discuss this zero-growth transition later.

The three-dimensional case is closer to reality but harder to
visualize.  The principles of our calculations here are identical to
those in 2D, with the particles able to move to any of 26 adjacent
sites.  The results are qualitatively similar, with the introduction
of a field reducing the fractal dimension without significantly
increasing density.  However, the effect of levellers is much more
pronounced: even a small number of molecules bringing about a
transition to a 100\% dense phase.  This might be anticipated from 
the existence of fluid percolation, which allows the molecules to 
move into the deposit, up to a much higher density.  However the
final 3D densities are significantly higher than the percolation threshhold: 
very close to 100\%.

\begin{figure}\protect{\includegraphics[width=\columnwidth]{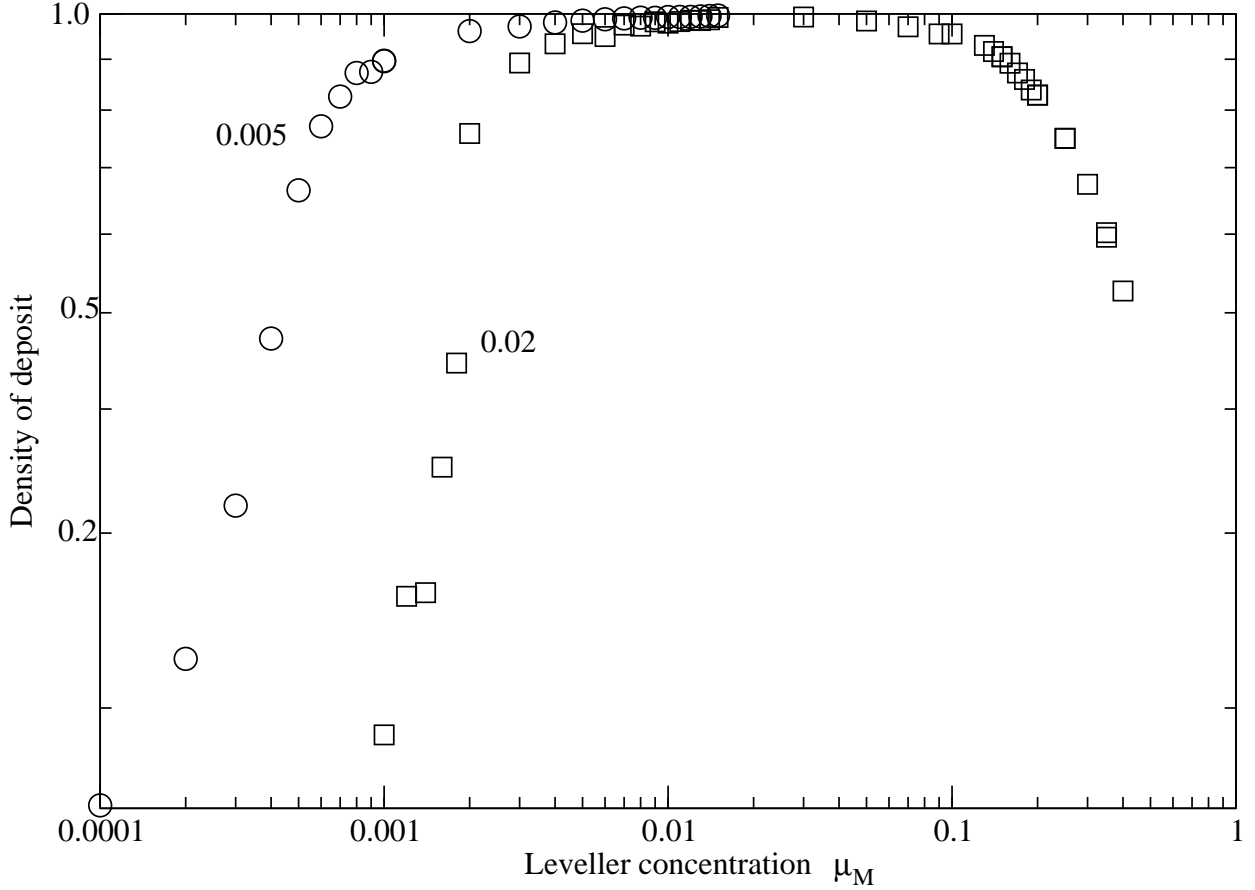}}
\caption{Logarithmic graph of density against concentration of
leveller for 3D model, with $\mu_I=0.005$(circles), $0.02$(squares),
$p=1$ and $F=0.2$.  Density is measured as the fraction of sites
occupied by an atom in the middle 80\% of the grid at the time when
the first atom appears in the top layer - this leads to a finite size
broadening of the transition.  Data is collected for a 100x200$^2$ grid.
The crossover to a dense deposit occurs for small leveller concentration, here 
when $\mu_M\approx\mu_I/10$, (it also depends on the field). 
Reduced density is observed again at very high leveller concentration: 
in this unphysical regime the leveller coats the surface, and is held 
in place by the downwards pressure of the falling ions, thus the reduced 
density comes from incorporated molecules, rather than an open structure. 
}\label{fig:3D}
\end{figure}

The complete compacting effect of the levellers for a typical example
is shown in figure \ref{fig:3D}.  A concentration of leveller
significantly less than the concentration of ions is required to obtain 
effectively 100\% density (with all the leveller molecules escaping).
Varying the field under these conditions has little effect - the sample remains dense.
For high $p\mu_M$, or for very low field, the deposition ceases.

Taken together the 2D and 3D results for this model of leveller action show
that while field or leveller molecules reduce the fractal dimension of
the deposit, only the levellers cause a significant densification of
the deposit.  Moreover, even a rather low concentration of molecules
with a low unsticking probability already has a strong densifying
effect.  In 2D it is not possible to produce a fully
dense deposit: the reason for this is that the large collapsing
dendrites fall and enclose regions which are then inaccessible to the 
leveller molecule.  In 3D the lower percolation threshhold leads to 
dense deposits.

The problem can be examined analytically in a mean field approach.
The continuity equation in the relevant regime, where the diffusion is 
controlled by the rate of unsticking (i.e. $D=\mu_Mp$), gives

\begin{equation}
\frac{d\rho_I(z)}{dt}= \rho_M(z)p \left ( \frac{d^2\rho_I(z)}{dz^2}-EC\frac{d\rho_I(z)}{dz} \right)
\end{equation}

where C is a geometric factor depending on the dimensionality and connectivity.

Since the molecules are unaffected by the ions, except for exclusion,  
$\rho_M(z)=\mu_M(1-\rho_I(Z)) $,  whence

\begin{equation}
\frac{d\rho_I(z)}{dt}= \mu_Mp(1-\rho_I)\left ( \frac{d^2\rho_I(z)}{dz^2}
-EC\frac{d\rho_I(z)}{dz} \right )
\end{equation}

with the boundary conditions that $\rho_I(0)=1$,
$\rho_I(\infty)=\mu_I$, the long time solution to this is
$\rho_I(z)=1$. The equation permits a steady state solution given by a
linear height/density profile $\frac{d\rho_I(z)}{dz}=-EC$, however the
boundary condition $\rho_I < 1$, makes this unphysical.  Thus there
are two solutions consistent with the boundary conditions:
either no deposition ($\rho_I=\mu_I$) or a dense deposit ($\rho_I=1$).

The mean field approach neglects the fact that only discrete numbers
of molecules are possible, and once excluded from a void unsticking
ceases inside that void.  The number of molecules incorporated
permanently in the deposit is rather small.  The unsticking ability of
the molecules allows them to escape, and here the value of unsticking
probability plays a role - low unsticking probability leads to extra
incorporation.  Thus although the mean field model tells us that the
deposit becomes 100\% dense independent of dimension, one should
recall that this assumption breaks down at high density.


In the mean field model, and in both 2D and 3D, there is a transition
from a growing state to a zero-growth-rate state.  This occurs when the
rate of removal of material from the deposit due to the action of the
levellers is equal to the rate of arrival. Growth on a fractal surface
occurs faster than on a dense one, and the effect of the levellers is
that the deposit grows and densifies.  The zero growth condition then
occurs for a dense deposit with a flat surface. The rate of arrival
per site is then simply $\mu_I$, while the escape rate is $p\mu_M$
times the probability that a released ion diffuses away.  For a flat
surface, this is approximately the probability that its first movement
is upwards, giving a rough estimate of the transition condition as
$\mu_I/\mu_M=9(1-E)p/26$.  In the simulations, we find that this type 
of zero-growth occurs only for low fields, $\mu_I$ and $\mu_M$. .  

For high fields, $\mu_I$ and $\mu_M$ crowding occurs, with the falling
ions dragging the levellers. This invalidates our mean field
approximation which assumes free diffusion. This results in a dense
mix of I, A and M autonomes at the bottom of the simulation, with
significant amounts of leveller incorporated in the deposit.  The atom
density in the deposit is therefore reduced (see fig. {\ref{fig:3D}.
The ion and leveller densities in this regime are far larger than
realized in experimental electrodeposition.

In figure \ref{fig:3D} where we consider varying $\mu_M$ 
with E fixed at 0.2  we see first the crossover from dendritic to dense
deposits and then the crossover to to non-growth region.


In summary, we have presented an extension of the diffusion limited
aggregation model which describes the unbinding effect of organic
molecules in microscopic detail.  The model shows that a very small
concentration of the levelling molecule acts to destroy the fractal
structure by a catalytic-type action which enables each molecule to
act multiple times.  In 3D the levellers have the most striking
effect, attaining deposit densities of over 99\% without themselves
being trapped.  Only a small amount of leveller is required to densify
the deposit, an intermediate amount speeds up the process, but beyond a
critical level all growth ceases.  Consistent with empirical practice 
in electrodeposition\cite{oniciu}, he optimum amount of leveller is 
somewhat less than the amount of ions, depending on the underlying 
lattice connectivity and the unsticking probability.

The action of the levelling molecules is most easily appreciated by
watching the system evolve dynamically.  A java applet which enables
the reader to do so in the 2D case can be seen at
www.ph.ed.ac.uk/nania/diffusion/diffusion.html


\end{document}